\documentclass[showpacs,amsmath,amssymb,twocolumn]{revtex4}
\usepackage[T1]{fontenc}
\usepackage{graphicx}
\usepackage{subfig}
\usepackage{epsfig}
\usepackage{dcolumn}
\usepackage{multirow}
\usepackage{rotating}
\usepackage{verbatim}
\usepackage{bm}
\usepackage{color}
\usepackage{soul}
\usepackage{float}
\usepackage{hyperref}

\def\lsim{\raise0.3ex\hbox{$<$\kern-0.75em\raise-1.1ex\hbox{$\sim$}}}
\def\gsim{\raise0.3ex\hbox{$>$\kern-0.75em\raise-1.1ex\hbox{$\sim$}}}

\def\pom{{I\!\!P}}

\def\beqa{\begin{eqnarray}}
\def\eeqa{\end{eqnarray}}

\begin{document}

\title{Investigating the diffractive gluon jet production in lepton-ion collisions}
%\pacs{12.38.Bx; 13.60.Hb;13.87.Ce;13.75.-Cs;13.85.Lg}
\author{G.M. Peccini} 
\email{guilherme.peccini@ufrgs.br}
\author{L. S. Moriggi}
\email{lucas.moriggi@ufrgs.br}
\author{ M.V.T. Machado}
\email{magnus@if.ufrgs.br}
\affiliation{High Energy Physics Phenomenology Group, GFPAE. Institute of Physics, Federal University of Rio Grande do Sul (UFRGS)\\
Caixa Postal 15051, CEP 91501-970, Porto Alegre, RS, Brazil}

\begin{abstract}

We study the diffractive jet production in electron-ion collisions in the kinematic region  where the mass $M_X$ of the diffractive final state is larger than $Q^2$. Based on parton saturation framework, predictions are done for the kinematics of future or possible $eA$ machines as the  EIC,  LHeC, HE-LHeC  and FCC-eA. We analyze the differential cross section as a function of jet (gluon) transverse momentum and from the experimental point of view this observable could be used to extract the saturation scale as a function of $x_{\pom}$. 
\end{abstract}

\maketitle
%\tableofcontents

\section{Introduction} 

The Electron - Ion Colliders (EIC) will open the possibility of probing the hadronic structure in the regime of large partonic densities and high strong field strengths, which are expected to modify the linear evolution equations. The EIC allows the measurement of inclusive and exclusive observables which are affected by the enhancement of non-linear effects in terms of the atomic mass number, $A$. In particular, within the parton saturation framework, the nuclear saturation scale, $Q_{s,A}$,  is enhanced with respect to the nucleon one, $Q_{s,p}$ by a sizable factor. For instance, for lead targets this enhancement of the nuclear saturation momentum reaches a factor  3  in contrast to the proton one where $Q_{s,p}(x=10^{-5})\approx 1$  GeV ($x$ is the usual Bjorken variable).  Particularly, in present study we consider  the simple ansatz proposed in Ref. \cite{Armesto:2004ud}, where the growth on atomic number $A$  depends on the quotient of the transverse parton densities to the power $1/\delta$,
\begin{eqnarray}
Q_{s,A}^2(x_{\pom};A)=Q_{s,p}^2 (x_{\pom})\left(\frac{A \pi R_p^2}{\pi R_A^2} \right ) ^{1/ \delta}
\label{qsnuc}
\end{eqnarray}
where $Q_{s,p}=(x_0/x)^{\lambda /2}$ GeV (parameters $x_0=4.2\times 10^{-5}$ and $\lambda=0.248$ are taken from the recent fit to high precision HERA data \cite{Golec-Biernat:2017lfv}) is the saturation scale of a single proton, $R_p$ is the proton radius and $R_A$ is the nucleus radius. For the latter quantity, we take the usual parametrization $R_A=(1.12 A^{1/3} - 0.86 A ^{-1/3})$. The quantities $\delta$ and $\pi R_p^2$ were  fitted \cite{Armesto:2004ud} from $\gamma A$ collisions at small-$x$ and their values are $0.79$ and $\pi R_p^2 = 1.55$  fm$^2$, respectively. Qualitatively, the nuclear saturation scale behaves like $Q_{s,A}^2\simeq A^{\Delta}Q_{s,p}^2$ with $\Delta \approx 4/9$. Quantitatively, for gold ($A=197$) and  lead nucleus ($A=208$) one gets $Q_{s,Au}^2\approx 2.8 \,Q_{s,p}^2$ and  $Q_{s,Pb}^2\approx 3\, Q_{s,p}^2$ , respectively. This very same ansatz enables to describe the $p_T$-integrated multiplicity in symmetric $AA$ collisions at mid-rapidity  \cite{Armesto:2004ud}. For processes probing perturbative typical scales like the photon virtuality $\mu^2=Q^2$ or $\mu^2=Q^2+m_V^2$ as in case of vector meson production an important part of observables are within the saturation region $\mu^2\lesssim Q_s^2$.  Recent studies have shown that the eA collider is the ideal facility to get deeper in the understanding of QCD at high energies \cite{Accardi:2012qut,Armesto:2019gxy}. In this context, the hard diffractive production is quite sensitive to unitarity corrections to the perturbative QCD calculation (pQCD). For example, in parton saturation regime contributions growing as $Q_s^2/Q^2$  are increasingly important and the leading-twist approximation of pQCD cannot account for such contributions.  A striking prediction of saturation approach is the constant ratio of the diffractive versus inclusive cross sections as observed at DESY-HERA as a function of photon-proton center-of-mass energy, $W_{\gamma^*p}$,  and the identification that diffractive deep inelastic scattering (DDIS)  is a semi-hard process \cite{GolecBiernat:1999qd}, that is, the  diffractive cross section is strongly sensitive to the infrared cutoff  given by $R_s=1/Q_s(x)$ and DDIS clearly  probes the transition region between the   dilute and  saturated regime. Furthermore, it has been shown that exclusive processes in $eA$ collisions can be nicely described using the geometric scaling property in parton saturation framework. In particular, the exclusive  light and heavy meson photonuclear production cross sections extracted from ultraperipheral heavy ion collisions are predicted without any further parameter fitting \cite{Ben:2017xny}.

In this work we investigate the gluon jet production in the diffractive photon dissociation in the context of the electron-ion colliders. In particular, we analyze the case for future electron-proton/nucleus colliders in the GeV regime (EIC) and in TeV regime as the  Large Hadron-electron Collider (LHeC) \cite{AbelleiraFernandez:2012cc}  as well as  the Future Circular Collider in electron-hadron mode (FCC-eh) \cite{Abada:2019lih}.  It will be considered the high diffractive mass $M_X$ kinematic region with the final state configuration, $e+p(A)\rightarrow e^{\prime}+ X+\mathrm{jet}+\mathrm{gap}+p(A)$,  having the (gluon) jet near to the edge of the rapidity gap. At the LHeC and FCC-eh the range of available momentum fraction of the diffractive exchange with respect to the proton can reach down to $x_{\pom}\simeq 10^{-5}$ for a large range of the momentum fraction of the parton  with respect to the diffractive exchange, $\beta = Q^2/(Q^2+M_X^2)$  (with $x=\beta x_{\pom}$). It was proposed in Ref. \cite{GolecBiernat:2005fe} that the measurement of the maximum of the differential cross section on the gluon (jet) transverse momentum $k_{\perp}$, i.e. $k_{\perp}^2 d^3\sigma_{\mathrm{diff}}^{\gamma}/d^2k_{\perp}dM_X$, provides a direct measurement of the saturation scale as a function of $x_{\pom}=(Q^2+M_X^2)/(Q^2+W^2)$. We will explore this possibility in what follows. Concerning the Electron Ion Collider (EIC), it seems to be very challenging to measure this type of events there once the kinematic reach for jet measurements at the EIC is found to be roughly $0.008 < x < 0.7$ and $ Q^2 > 25$ GeV$^2$ for $\sqrt{s}= 89$ GeV \cite{Arratia:2019vju}. Nevertheless, it is more likely to perform this measurements at high energy machines (LHeC and FCC) or in ultraperipheral  $AA$ collisions with a rich content of quasi-real photons at the LHC.

This study can be complementary to recent investigations of diffractive dijet production in $\gamma^*h$ collisions (with $h=p,A$). In what follows, we summarize the main studies along this direction. The exclusive dijet production is investigated in \cite{Altinoluk:2015dpi} within the Color Glass Condensate (CGC) formalism at leading order (LO) demonstrating that the azimuthal angle correlations and momentum transfer, $t$, distributions are sensitive to parton saturation at small-$x$. Important points are the relation between the increasing of saturation scale, $Q_{s,A}$, and  an enhancement of away-side correlations as well as the present of dips in $t$-dependence which is absent on non-saturation models. In Ref. \cite{ Hatta:2016dxp} the angular correlation between the transverse momentum of the produced dijet and the recoiled momentum of the nucleon is investigated in the context of the quantum phase space of Wigner distribution of small-$x$ partons. It was pointed out that the gluon Wigner distributions are closely related to the impact parameter dependent dipole and quadrupole scattering amplitudes and they could be measured in diffractive DIS in $eA$ collisions at an EIC or in ultraperipheral collisions at the LHC. The last case was addressed using NLO pQCD  in Refs. \cite{Guzey:2016tek,Guzey:2018dlm} for both diffractive and inclusive dijet production. Similarly, in Ref. \cite{Mantysaari:2019csc} the  gluon in Wigner and Husimi distributions of nucleons were considered
   within the CGC formalism including numerical solution of the JIMWLK equations. The anisotropy of these distributions as a function of the angle between impact parameter and transverse momentum has been analyzed and  signatures of these angular correlations were proposed for EICs.  Along the same lines, taking into account the multi-gluon correlations inside nuclear targets at small-$x$ in Ref. \cite{Mantysaari:2019hkq} the elliptic modulation of diffractive dijets  was investigated  and it was shown that saturation effects are significant when looking at the nuclear modification of the ratio between the differential inclusive and diffractive dijet cross sections. Authors of \cite{Hatta:2019ixj} studied the soft gluon radiation associated with the final state jets and an all order resummation formula has been derived. They argued that soft gluon resummation plays an important role in EIC and helps to explore the nucleus tomography. The impact parameter dependence was  studied analytically (including elliptic anisotropy)  for  coherent diffractive dijet production in $ep$ and $eA$ collisions in Ref. \cite{Salazar:2019ncp}. General relations are found connecting angular correlations of the dipole orientation and $b$-vector in coordinate space with angular correlations between mean dijet $k_{\perp}$ and hadron recoil momentum. Finally, from theoretical point of view a  complete NLO description of diffractive dijet production is carried out in Ref. \cite{Boussarie:2019ero}, where the direct coupling of the Pomeron (viewed as a color singlet QCD shock wave)  to the diffractive $X$ state is considered. The numerical results are promising mostly at intermediate to large $\beta$ values.

This paper is organized as follows: in the next section, we determine the expression for the differential cross section for diffractive gluon jet in terms of the transverse momentum scale and the diffractive mass. Afterwards, we show the results applying this formalism taking into account the expected kinematic-plane for the planned high energy  lepton-ion machines, as well as considering different diffractive masses. The feasibility of extracting the saturation scale from measured cross section will be carried out.  Finally, we conclude by summarizing the main ideas that can be extracted from the presented results.

\section{Theoretical Framework}

The dipole approach is a convenient way to calculate observables at high energies, such as the total and diffrative cross sections once the dipole picture makes possible the factorization of the whole process, which in turn is divided in a QED (the photon fluctuating into a quark-antiquark pair) and QCD (the interaction between the dipole and the hadron) subprocesses. Such a mechanism is only possible due to the fact that the time of fluctuation of the photon is much bigger than the time of interaction between the dipole and the target at high energies.  
In this formalism, the photon fluctuates into a quark-antiquark pair of transverse size $r \sim  1/Q$, where $Q^2$ is the photon virtuality. The wave functions corresponding to the photon (with transverse and longitudinal polarizations)  fluctuating into this pair are taken from the light cone perturbative theory, and are given by
\begin{eqnarray}
|\Psi_{T}(z,\vec{r},Q^2)|^2 & = &\frac{6 \alpha_{em}}{4 \pi ^2}\sum _{f} e_f^2[z^2+(1-z)^2] \epsilon ^2 K_1^2(\epsilon r) \nonumber \\
&+& m_f^2K_0(\epsilon r), \\
|\Psi_{L}(z,\vec{r},Q^2)|^2 &=&\frac{6 \alpha_{em}}{\pi ^2} \sum _{f} e_f^2 [Q^2 z^2 (1-z)^2 K_0^2 (\epsilon r)],
\label{photonwfs}
\end{eqnarray}
where $\psi_{T}$ stands for the transverse part of the photon wave function, whereas $\psi_{L}$ is its longitudinal contribution. The quantity $\vec{r}$ is the relative transverse separation between the quark and the antiquark and $z (1-z)$ is the longitudinal momentum fraction of the quark (antiquark) whose flavor is $f$.  Also in this picture, the total and diffractive cross sections can be calculated as follows (sum of flavors is implicit),
\begin{eqnarray}
\sigma_{tot}^{\gamma^*p}(x,Q^2)&=& \int dz d^2\vec{r} \left( |\psi_{T}|^2 +|\psi_{L}|^2 \right)\,2 \int d^2\vec{b} N(x,\vec{r},\vec{b}) , \nonumber \\
\sigma_{\mathrm{diff}}^{\gamma^*p}(x,Q^2)&=& \int dz d^2\vec{r} \left( |\psi_{T}|^2 +|\psi_{L}|^2 \right) \int d^2\vec{b} |N(x,\vec{r},\vec{b})|^2, \nonumber
\end{eqnarray}
where $N(x,\vec{r},\vec{b})$ is the dipole scattering amplitude for QCD color dipoles having transverse sizes $\vec{r}$ at impact  parameter $\vec{b}$ and probing Bjorken-$x$ variable in the target. The dipole amplitude is related to the $S$-matrix, with $S(x,\vec{r},\vec{b})= 1-N(x,\vec{r},\vec{b})$. In the expressions above, the variable $\epsilon$ is defined as $\epsilon=\sqrt{z(1-z)Q^2 +m_f^2}$, where $m_f$ is the quark mass of flavour $f$. For simplicity, in this work we will only consider light quarks ($u$, $d$, $s$) with masses $m_f=0.14 \ GeV$. The quantities $K_0$ and $K_1$ are the the Modified Bessel Functions of Second Kind of order zero and one, respectively.

In the present paper we aim to analyze the diffractive gluon-jet production in diffractive dissociation of photons in DIS, investigating the nuclear effects when taking into account nuclei as targets. This is relevant for the physics to be studied in EIC and LHeC/FCC-eh machines. The starting point is to write the diffractive cross section in terms of the decomposition on the Fock states of incident virtual photon, $|\gamma^*\rangle = |q\bar{q}\rangle +  |q\bar{q}g\rangle  \ldots$, where the $q\bar{q}$ colorless dipole is characterized by the wavefunctions in Eq. (\ref{photonwfs}). The second Fock state includes the  emission of a soft gluon (small longitudinal momentum fraction, $z_g$) off a $q\bar{q}$ dipole and its transverse momentum can be identified with the  momentum of the jet closest to the rapidity gap. We are interested in this last component, which is dominant in the kinematic regime where the diffractive mass, $M_X$, is larger than the photon virtuality ($M_X^2\gg Q^2$). The terms  from jets initiated by quarks in such a kinematic interval are suppressed. In the Pomeron language, this corresponds to a momentum fraction of the parton with respect to the diffractive exchange having $\beta \ll1$. In Ref. \cite{GolecBiernat:2005fe} the diffractive cross section for the production of a gluon having transverse momentum $k_{\perp}$ and rapidity $y$ on the collision of a $q\bar{q}$ of transverse size $r$ with the target has been derived. The relevant diagrams include the cases where the interaction with the target  takes place after and before the gluon emission. The corresponding differential cross section in leading $\ln(1/\beta)$ accuracy and small $Q^2$ is given by \cite{GolecBiernat:2005fe},
\begin{eqnarray}
\frac{d\sigma_{\mathrm{diff}}^{q\bar{q}g}}{d^2k_{\perp}dM_X}&=& \frac{2M_X}{Q^2+M_X^2}\int d^2\vec{r}d^2\vec{b}\, \rho (r,Q^2)\frac{d\sigma_g(\vec{r},\vec{b})}{d^2k_{\perp}dy},
\label{diffxs}
\end{eqnarray}
\begin{eqnarray}
\frac{d\sigma_g(\vec{r},\vec{b})}{d^2k_{\perp}dy}&=& \frac{\alpha_sN_c^2}{4\pi^2C_F}\,A(k_{\perp}, x_{0,1};\Delta \eta)A^*(k_{\perp}, x_{0,1};\Delta \eta), \nonumber
\end{eqnarray}
where $\rho (r,Q^2)=\int dz (|\psi _{T}^{\gamma} (r,z;Q^2)|^2+|\psi_{L}^{\gamma}(r,z;Q^2)|^2)$ and $x_{0,1}=b\pm (r/2)$ ($x_0$ and $x_1$ are the transverse positions of $q$ and $\bar{q}$, respectively).  The rapidity gap is written as $\Delta \eta = \log (1/x_{\pom})=Y-y$  with $Y=\log (1/x)$ being the total rapidity. The quantity $A(k_{\perp}, x_0,x_1; \Delta \eta)$ is written \cite{GolecBiernat:2005fe} in terms of the elastic $S$-matrix for the collisions of the dipole on the target evolved at the rapidity $\Delta \eta$, $S(x_0,x_1;\Delta \eta)$, and the elastic $S$-matrix for the collision of two dipoles, $S^{(2)}(X_0,x_g,x_1;\Delta \eta)$, where $x_g$ is the gluon transverse coordinate. Independently of the specific form for $S$-matrices the quantity $k_{\perp}^2d\sigma/d^2k_{\perp}dM_X$ rises as $k_{\perp}^2$ for small gluon transverse momenta whereas falls as $1/k_{\perp}^2$ for large ones. A maximum occurs for a typical transverse momentum where parton saturation becomes important, i.e., $(k_{\perp})_{max}\propto Q_s$ where $Q_s(x_{\pom})$ is the saturation scale. 

In Ref.  \cite{GolecBiernat:2005fe} a simplified model for the $S$-matrices has been considered. Inspired in the GBW model \cite{GolecBiernat:1999qd} and neglecting correlations between the  two dipoles in $S^{(2)}$, they read as,
\begin{eqnarray}
S(x_0,x_1;\Delta \eta) &=&  e^{-\frac{(Q_sr)^2}{4}}\Theta (R-|b|) + \Theta (|b| -R),\nonumber \\
S^{(2)}(x_0,x_1, x_g;\Delta \eta) &=&  e^{-\frac{Q_s^2[(x_0-x_g)^2+((x_g-x_1)^2]}{4}}\Theta (R-|b|)\nonumber \\
 &+&  \Theta (|b| -R),
 \label{gbwss}
\end{eqnarray}
where $R$ is the target radius and the saturation scale depends on $x_{\pom}$ variable.  The theta function appearing in $S$-matrices will give an overall normalization factor after $b$-integration in Eq. (\ref{diffxs}) in the form $\bar{\sigma}_0=\pi R^2$. The parameter $\sigma_0=2\pi R^2=2\bar{\sigma }_0=27.32$ mb for proton target has been fitted from DESY-HERA data on proton structure functions at small-$x$  \cite{GolecBiernat:1999qd}. In Ref. \cite{Marquet:2007nf} a different model for the $S$-matrices has been considered, where their impact parameter dependence was factorized having a profile in the form $T(b)=e^{-b^2/(2B_D)}$, where $B_D\simeq 6$ GeV$^{-2}$ is the diffractive slope and  $\sigma_0=4\pi B_D$. Moreover, the $S^{(2)}$ is expressed in terms of color dipole amplitude $N(r;x_{\pom})$ taken fro Iancu-Itakura-Munier (IIM) \cite{Iancu:2003ge} saturation model (with $S=1-N$). In particular, in the small-$\beta$ limit it was considered, $N^{(2)}(x_0,x_1,x_g, \Delta \eta)= N(|\vec{x}_0-\vec{x}_g|Q_s, \Delta \eta)+  N(|\vec{x}_g-\vec{x}_1|Q_s, \Delta \eta)- N(|\vec{x}_0-\vec{x}_g|Q_s|, \Delta \eta)N(\vec{x}_g-\vec{x}_1|Q_s, \Delta \eta)$.

Taking into account the GBW-like parametrization, Eqs. (\ref{gbwss}), the integration over impact parameter in Eq. (\ref{diffxs}) can be done. That model contains the main features which are also present in more sophisticated models for the dipole amplitude. This will give a semi-analytical expression for the differential cross section (with $|\vec{k}_{\perp}|=\kappa$),
\begin{eqnarray}
\frac{d\sigma_{\mathrm{diff}}}{d^2 k_{\perp}d M_X} & = &  \frac{\alpha_s N_c^2 \bar{\sigma_0}} { 4 \pi ^2 C_F} \frac{M_X}{M_X^2 +Q^2}  \int dr^2 d\theta\, \rho (r,Q^2) \nonumber \\
&\times& \left(\frac{e ^{-r^2 Q_s^2 / 2}}{\kappa^2}\right)\frac{1}{\left[\frac{\kappa}{(r Q_s^2)} -\frac{r Q_s^2}{4\kappa}\right]^2 +\cos^2 \theta }\nonumber \\
&\times & \left[ T_1(r,\kappa,Q_s) +   T_2(r,\kappa,Q_s) +  T_3(r,\kappa,Q_s)\right], \nonumber \\
\label{semianxs}
\end{eqnarray}
where the auxiliary functions $T_{1,2,3}$ are given by,
\begin{eqnarray}
T_1 & = &  \left[ \cos \left(\frac{1}{2} \kappa r \cos \theta \right) -e ^{-\kappa^2 / (2Q_s^2) + Q_s^2 r^2 / 8} \right]^2,  \\
T_2 &=& \frac{Q_s^4 r^2}{4\kappa^2} \sin^2\left(\frac{1}{2} \kappa r \cos \theta \right), \\
T_3 &=&  \frac{rQ_s^2}{\kappa} \cos \theta \sin \left(\frac{1}{2} \kappa r \cos \theta \right) \nonumber \\
&\times& \left[ \cos \left(\frac{1}{2} \kappa r \cos \theta \right) -e ^{-\kappa^2 / (2Q_s^2)+Q_s^2 r^2 / 8 } \right].\\
\end{eqnarray}

Before computing numerically the cross section above, it would be interesting to investigate its qualitative behavior. It is well known that the virtual photon overlap function times dipole transverse size, $r\rho (r,Q^2)$, presents a peak at $r\simeq d/Q$ (with $d\approx 2$). Moreover, in the region studied here, $M_X^2\gg Q^2$, the prefactor $M_X^2/(M_X^2+Q^2)\rightarrow 1$. Taking into account an angle average cross section, Eq. (\ref{semianxs}), simplifies to,
\begin{eqnarray}
\langle k_{\perp}^2\frac{d\sigma_{\mathrm{diff}}}{d^2 k_{\perp}d M_X}  \rangle &  \propto & \frac{e ^{-d^2Q_s^2 / 2Q^2}}{\left[\frac{\kappa Q}{(d Q_s^2)} -\frac{d Q_s^2/Q^2}{4\kappa}\right]^2 + \frac{1}{2} } \nonumber \\
&\times &\left[\frac{1}{2} - e^{ -\frac{\kappa^2}{Q_s^2}+\frac{d^2Q_s^2}{4Q^2}} + \frac{Q_s^4 d^2}{4Q^2\kappa^2} \right],
\end{eqnarray}
that for the case of $Q^2\gg Q_s^2$ and assuming $d=2$ gives the qualitative behavior,
\begin{eqnarray}
\langle k_{\perp}^2\frac{d\sigma_{\mathrm{diff}}}{d^2 k_{\perp}d M_X}  \rangle  \propto  \frac{\frac{1}{2} - e^{ -(\frac{\kappa^2}{Q_s^2})} + (\frac{Q_s^2}{Q^2})(\frac{Q_s^2}{\kappa^2})}{\left[\frac{1}{2}(\frac{\kappa}{Q_s})(\frac{ Q}{ Q_s}) -\frac{1}{2}(\frac{ Q_s}{\kappa})(\frac{Q_s}{ Q})\right]^2 + \frac{1}{2} }, 
\end{eqnarray}
which is a function dependent on the ratios $\kappa/Q_s$ and $Q/Q_s$. For a fixed $Q^2$, for large $\kappa \gg Q_s$ the differential cross section falls as $1/\kappa^2$. 

To avoid the uncertainties concerning the running coupling $\alpha_s$ and the parameter $\sigma_0$ (which comes from the GBW parametrization, see \cite{Golec-Biernat:2017lfv} for recent analyzes), the following quantity is defined,
\begin{eqnarray}
\sigma^{\mathrm{scaled}} (\kappa,Q^2,Q_s)=\frac{1}{\alpha_s \sigma_0} \left( \frac{M_X^2+Q^2}{M_X^2} \right) M_X \frac{d \sigma _{\mathrm{diff}}}{d^2k_{\perp} dM_X}.
\label{scaledxs}
\end{eqnarray}

\begin{figure}
    \includegraphics[width=\columnwidth]{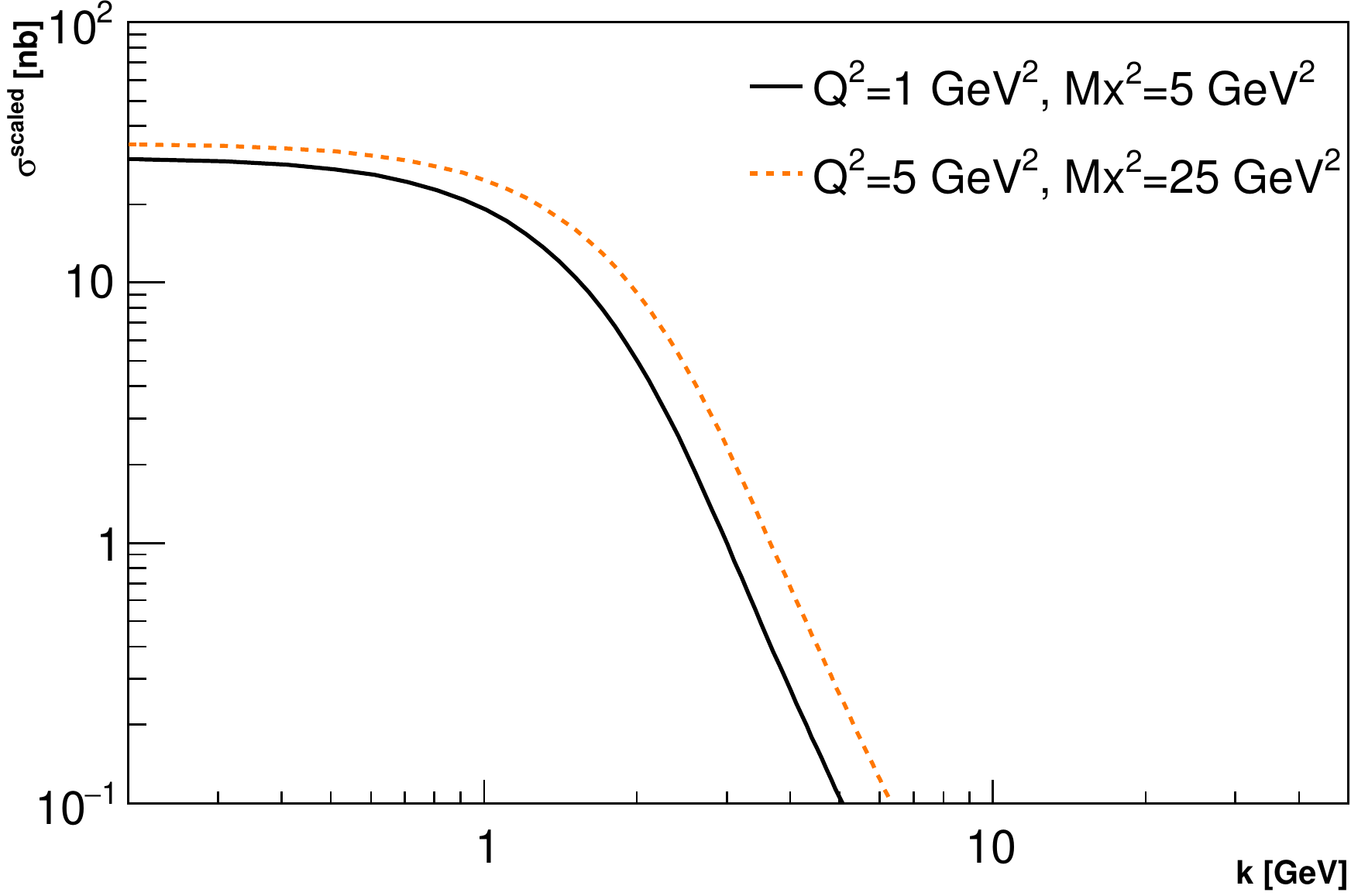}
    \caption{Diffractive jet (gluon) production at EIC  ($\sqrt{s}=92$ GeV) as a function of jet transverse momentum $\kappa$ for the configurations ($Q^2=1$ GeV$^2$, $M_X^2=5$ GeV$^2$) and ($Q^2=5$ GeV$^2$, $M_X^2=25$ GeV$^2$).}
    \label{fig:1}
\end{figure}

Let us now perform the corresponding phenomenology for diffractive gluon jet production in the context of electron-nucleus collisions. For the  saturation scale for protons we consider the usual power-like behavior, $Q_{s,p}(x_{\pom})=(x_0/x_{\pom})^{\lambda/2}$  GeV. The parameters $\lambda$ and $x_0$ were taken by fitting HERA data and their values are $\lambda=0.248$ and $x_0=4.2 \times 10^{-5}$, respectively \cite{Golec-Biernat:2017lfv}. The variable $x_{\pom}$ represents the longitudinal momentum fraction carried by the Pomeron, which is the exchanged object in diffractive processes.  In order to compute the nuclear saturation scale $Q_{s,A}$, we take the simple ansatz proposed in Ref. \cite{Armesto:2004ud}, as presented in Eq. (\ref{qsnuc}) in the introduction section. We have shown that  $Q_{s,Au}^2\approx 2.8 \,Q_{s,p}^2$ and  $Q_{s,Pb}^2\approx 3 \,Q_{s,p}^2$ , respectively.  Notice that the the value of the nuclear saturation scale can vary whether distinct treatments of the nuclear collision geometry are considered. For instance, using a local saturation scale, $Q_s^2(x,b)=Q_s^2(x,b=0)T_A(b)$ with $T_A$ being the nuclear thickness function, and a Gaussian $b$-profile the relation between $Q_{s,A}$ and $Q_{s,p}$ is found \cite{Salazar:2019ncp}. In the hard sphere approximation for the nuclear density $\rho_A$, we have $Q_{s,A}^2=3A(R_p/R_A)^2Q_{s,p}^2$. This will give $Q_{s,Au}^2\approx 2.2 Q_{s,p}^2$ and  $Q_{s,Pb}^2\approx 2.3 Q_{s,p}^2$. Thus, typically the theoretical uncertainty on the determination of the saturation scale compared to the proton one is of order 20$\%$. Accordingly, in nuclear case the overall normalization will be replaced by $\sigma_0\rightarrow \sigma_A=2\pi R_A^2$.

A different prescription for introducing nuclear effects can be used as writing down the $S$-matrices in terms of a Glauber model for the dipole-nucleus cross section, $N_A(x,r,b)$, using the model in Ref. \cite{Armesto:2002ny} for instance. Another possibility is to consider the recently determined dipole amplitude depending on impact parameter determined from numerical solution of the  Balitsky-Kovchegov (BK) equation with the collinearly improved kernel \cite{Cepila:2020xol}. Eventually, it can be considered also the model of the proton as constituted by hot spots (representing regions of high gluon density), where  its structure changes from interaction to interaction. This idea has been successfully applied for exclusive photonuclear production of vector mesons in Refs. \cite{Cepila:2017nef,Bendova:2018bbb}. In the next section we apply the geometric scaling ansatz for obtaining estimates of differential cross section as a function of gluon transverse momentum for planned electron-ion machines bearing in mind the theoretical uncertainties in $S$-matrix in the nuclear case.

\section{Results and Discussions}
\begin{figure*}[t]
\subfloat[]{
\includegraphics[height=6cm]{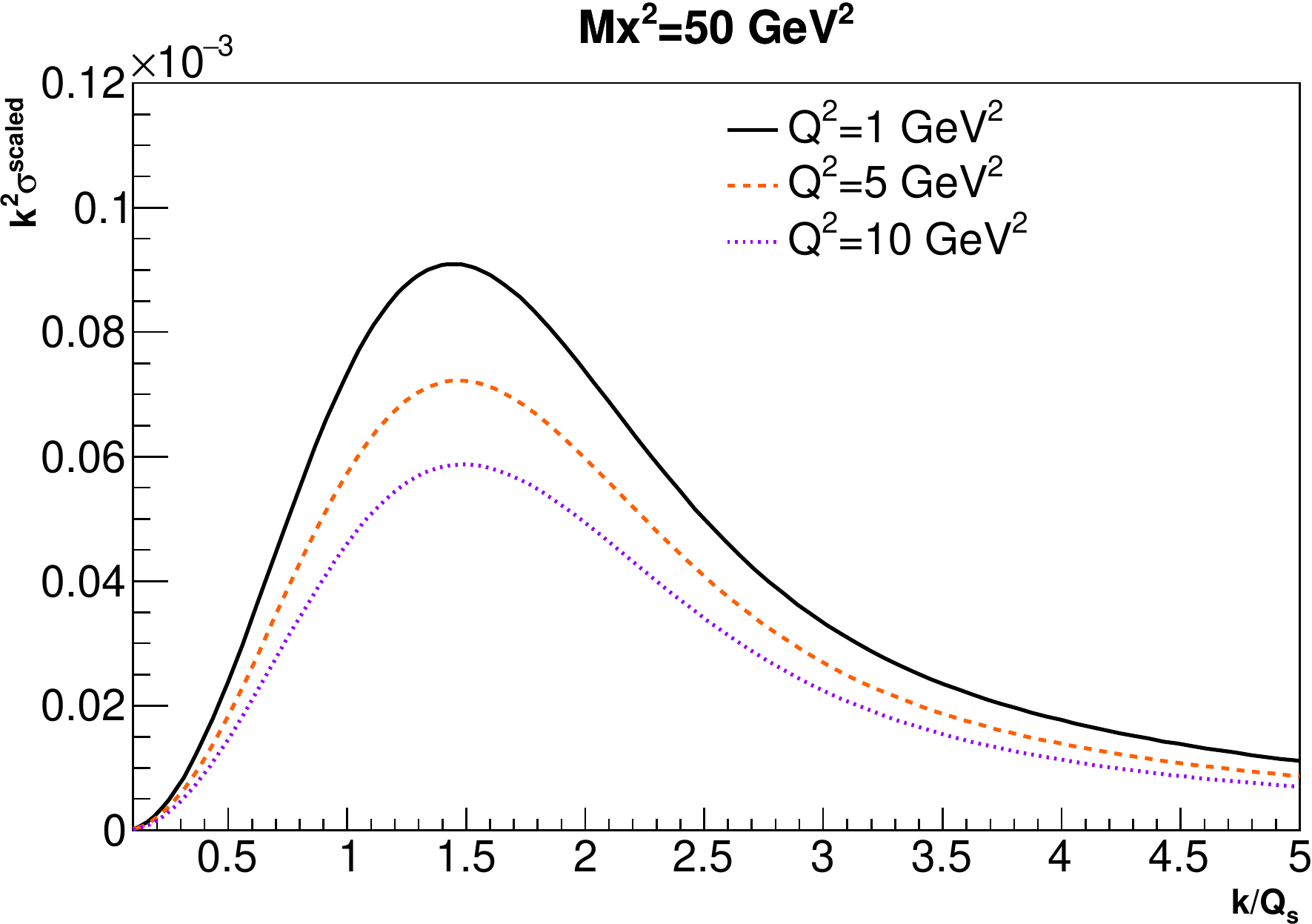}
\label{fig2a}
}
\quad %espaco separador
\subfloat[]{
\includegraphics[height=6cm]{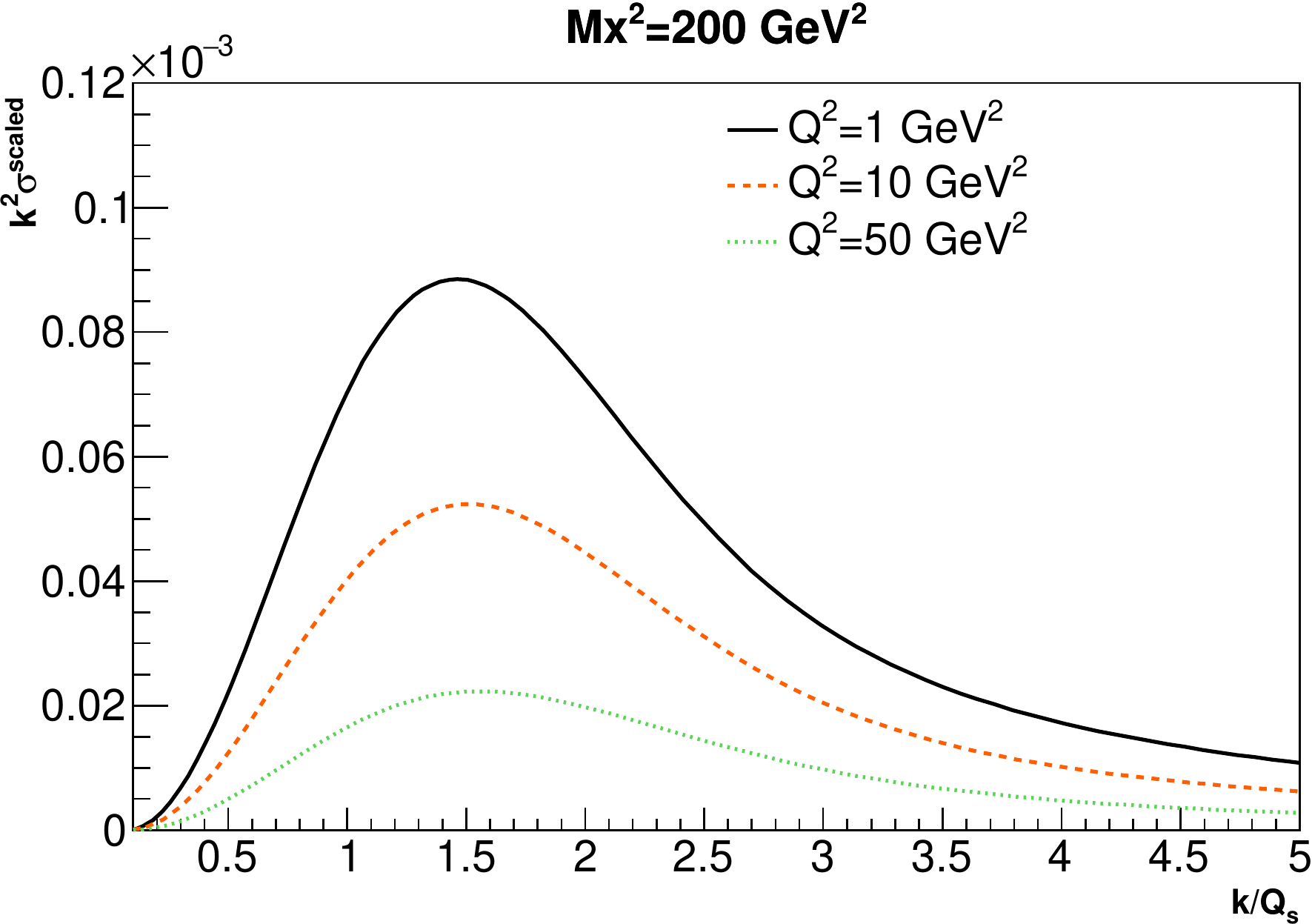}
\label{fig2b}
}
\caption{Differential cross section $\kappa^2\sigma^{\mathrm{scaled}}$  at the LHeC  ($\sqrt{s}=812$ GeV) as a function of  $\tau=(\kappa/Q_s)$. The following configurations are shown: (a) for $M_X^2 = 50$ GeV$^2$ with $Q^2=1,\,5,\,10$ GeV$^2$  and (b) for $M_X^2 = 200$ GeV$^2$ with $Q^2=1,\,10,\,50$ GeV$^2$ . The peak occurs around  $\tau_A \approx1.5$.}
\label{fig:2}
\end{figure*}

In this section, we numerically evaluate the formula for the gluon jet differential cross section, Eqs. (\ref{semianxs}) and (\ref{scaledxs}),  using the nuclear saturation scale based on geometric scaling property,  Eq. (\ref{qsnuc}). We perform our analysis for diffraction in $eA$ collisions focusing only on  coherent diffraction $e +A \rightarrow e + X + A$, where the incident nucleus remains intact in the final state. Incoherent diffraction, $e +A \rightarrow e + X + A^*$, which dominates for large-$|t|$ is out of scope of the present study.  We summarize in Table \ref{tab:1} the investigated energy configurations (in units of GeV) of planned electron-ion colliders, where $\sqrt{s}$  is the center-of-mass collision energy per nucleon and $xys=Q^2$ ($y$ is the inelasticity variable).

\begin{table}[t]
\caption{The design center-of-mass energy (in unities of GeV) for electron-nucleus collisions in the  machines EIC, LHeC, high energy upgrade of LHeC (HE-LHeC) and FCC-eA, respectively.}
\begin{center}
\begin{tabular}{|l|c|c|c|}
\hline 
Collider & $E_e$ & $E_A$ & $\sqrt{s}$ \\
\hline
EIC     & 21 & 100  &  92 \\
LHeC    & 60 & 2760 &  812 \\
HE-LHeC & 60 & 4930 &  1088 \\
FCC-eA     & 60 & 19700 &  2174 \\
\hline
\end{tabular}
\end{center}
\label{tab:1}
\end{table}

We start the analyses for the EIC \cite{Accardi:2012qut}, presenting the scaled cross section as a function of jet transverse momentum, $\kappa$.   For a gold nucleus, in Fig. \ref{fig:1} the results are shown for the scaled cross section, Eq. (\ref{scaledxs}), in the following two  kinematic configurations: $Q^2=1$ GeV$^2$ and $M_X^2=5$ GeV$^2$ (solid line) and $Q^2=5$ GeV$^2$ and $M_X^2=25$ GeV$^2$ (dashed line).  These values correspond to ($\beta \simeq 0.17,\, x_{\pom}\simeq 7.0\times 10^{-4}$) and ($\beta \simeq 0.17,\, x_{\pom}\simeq 3.5\times 10^{-3}$), respectively. The rapidity gap is $\Delta \eta \approx 3$ and the more prominent feature is the plateau for $\kappa \lesssim1$ GeV.  This feature is also observed  in $ep$ case \cite{GolecBiernat:2005fe}  and explained by the fact that the differential cross section $\kappa^2d\sigma/d^2k_{\perp}dM_X$ rises as $\kappa^2$   for small transverse momentum as referred already.  This happens independently of the particular model for the $S$-matrices.  On the other hand, at relative large $\kappa$, the cross section falls as $1/\kappa^4$ and the transition region is driven by the nuclear saturation scale, $Q_{s,Au}^2(x_{\pom}\sim 10^{-3})\approx1.3$ GeV$^2$.  By using $\sigma_A\simeq 267$ fm$^2$ and $\alpha_s=0.2$ we estimate the following values for the differential cross section at $\kappa = 10$ GeV:
\begin{eqnarray}
M_X\frac{d\sigma_{\mathrm{diff}}}{d^2k_{\perp}dM_X} &\approx & 8 \,\frac{nb}{GeV^2} ,\,Q^2=1\,GeV^2, M_X^2=5\, GeV^2,\nonumber \\
M_X\frac{d\sigma_{\mathrm{diff}}}{d^2k_{\perp}dM_X} & \approx & 17 \,\frac{nb}{GeV^2},\,Q^2=5\, GeV^2, M_X^2=25\, GeV^2,\nonumber 
\end{eqnarray} 
\begin{figure*}[t]
\subfloat[]{
\includegraphics[height=5.9cm]{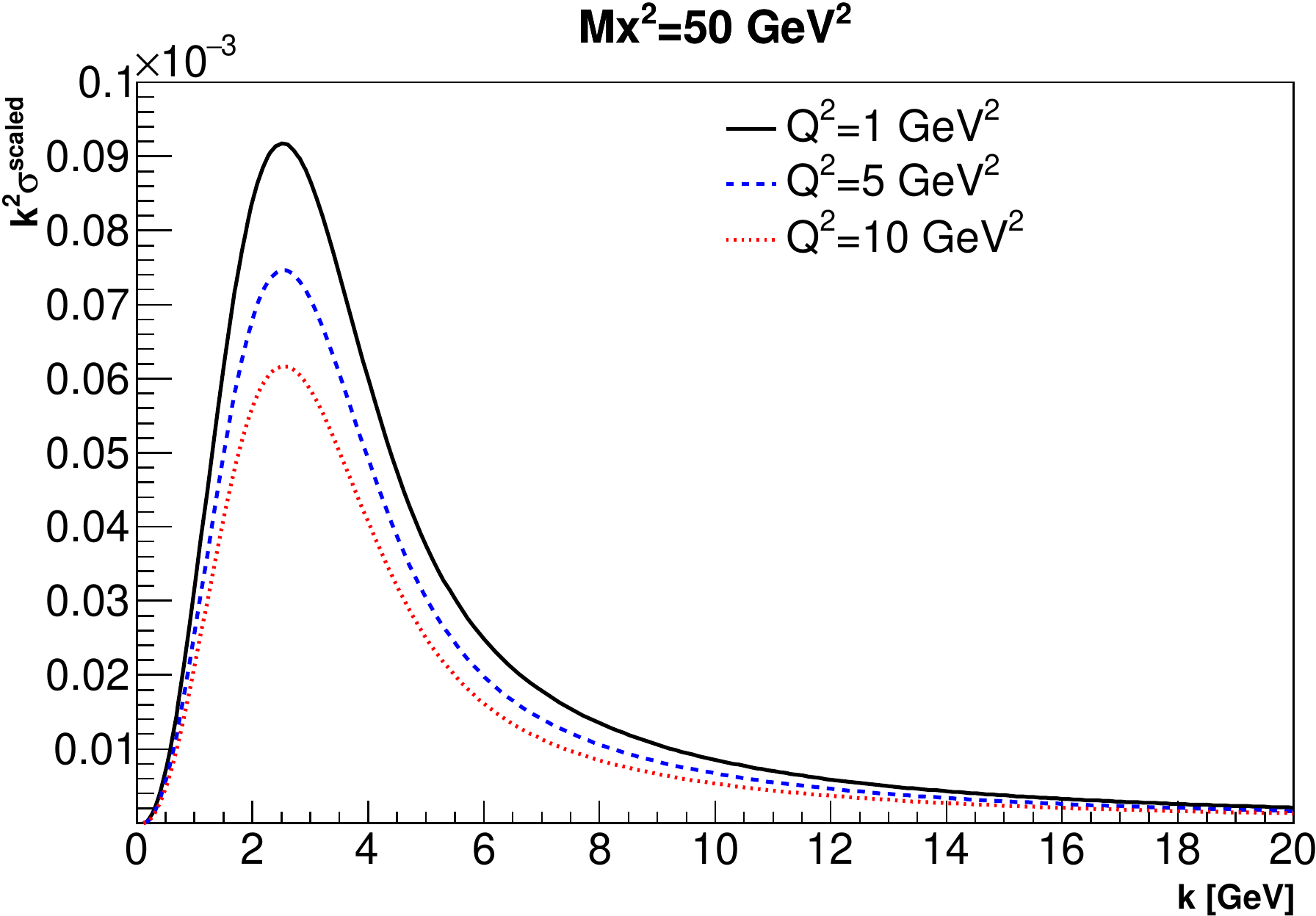}
\label{fig3a}
}
\quad %espaco separador
\subfloat[]{
\includegraphics[height=5.9cm]{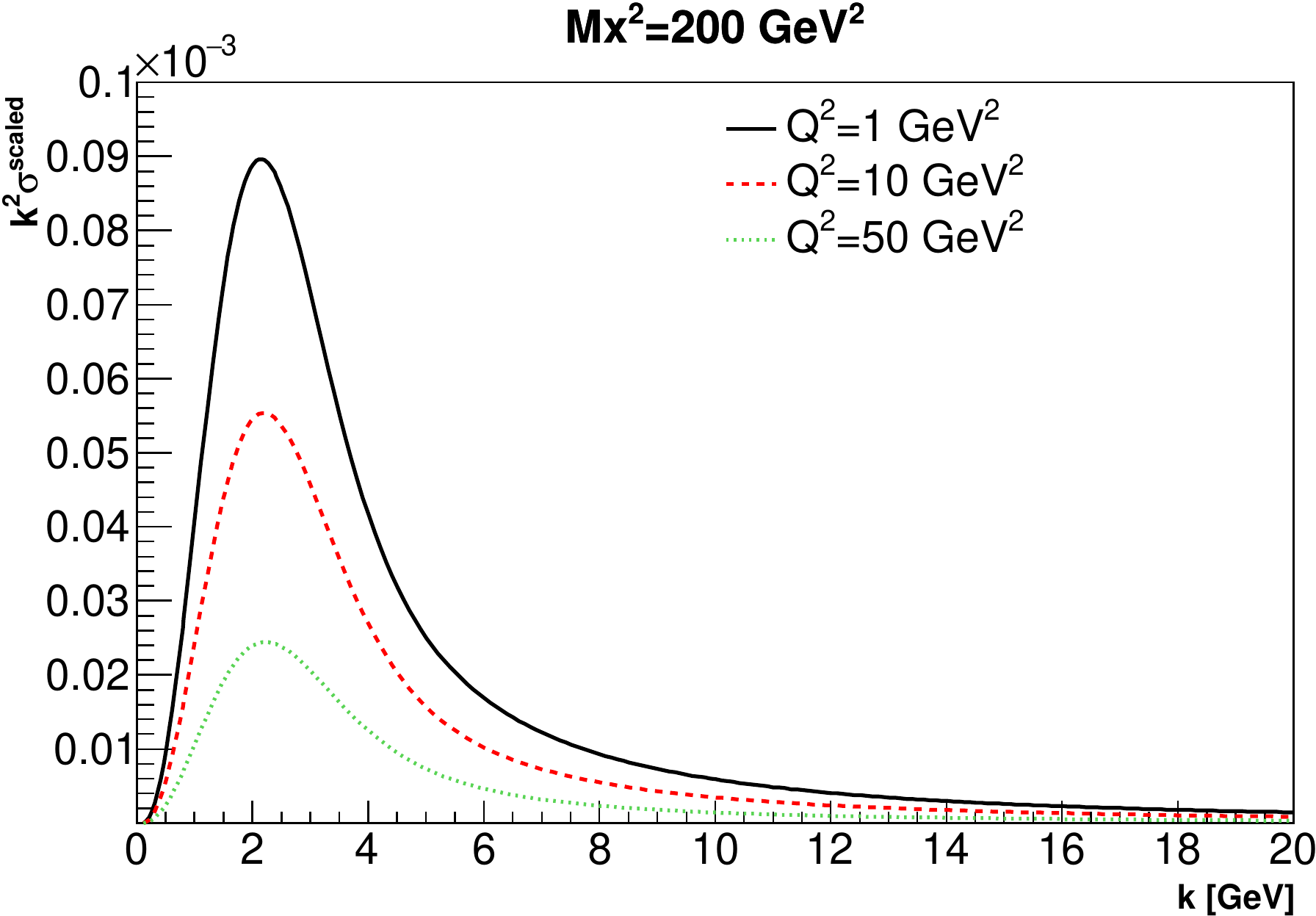}
\label{fig3b}
}
\caption{Differential cross section $\kappa^2\sigma^{\mathrm{scaled}}$  at the HE-LHeC  ($\sqrt{s}=1.088$ TeV) as a function of  $\kappa$. Three  configurations are shown: $M_X^2 = 50$ GeV$^2$ with $Q^2=1,\,5,\,10$ GeV$^2$  and $M_X^2 = 200$ GeV$^2$ with $Q^2=1,\,10,\,50$ GeV$^2$ .}
\label{fig:3}
\end{figure*}

We now turn to the LHeC in its heavy-ion mode \cite{Bruning:2019scy}, which would scatter electrons with $E_e=60$ GeV on a beam of nuclei from the LHC, with $E_A=2.75$ TeV per nucleon resulting in $\sqrt{s}=812$ GeV per nucleon. The corresponding  integrated luminosity could reach 10 fb$^{-1}$, being $10\times$ bigger than the full integrated luminosity achieved in $ep$ collisions at DESY-HERA. Due to the high luminosity, the LHeC or equivalent high energy machine opens the opportunity to directly measure the nuclear saturation scale as a function of $x_{\pom}$ as firstly proposed in  \cite{GolecBiernat:2005fe}. Specifically, whether the cross section  $\kappa^2d\sigma/d^2k_{\perp}dM_X$  can be measured as a function of $\kappa$ for distinct values of $x_{\pom}$ the positions of its maximum is translated into the $x_{\pom}$-dependence of saturation scale. Using the same reasoning the absolute value of $Q_{s,A}$   could be determined by considering a wide interval of $Q^2$ in the limit $\beta \ll 1$.  The property is showed in Fig. \ref{fig:2}, where the cross section $\kappa^2\ \sigma^{\mathrm{scaled}}(\kappa,Q^2,Q_{s,Pb}^2)$ is presented as a function of the jet transverse momentum scaled by the nuclear saturation scale, $k/Q_{s,A}(x_{\pom})$. To quantify the dependence of the position of the bump, we plot the cross section for 3  distinct values of photon virtuality and it can be clearly seen that the location of bumps do not depend on $Q^2$ at all. It  is straightforward to notice the marked bumps that separate the saturation region from the linear one. The numerical results are for (a) $M_X^2=50$ GeV$^2$  at virtualities $Q^2=1$ GeV$^2$ (solid  line), $Q^2=5$ GeV$^2$ (dashed line) and (b) $Q^2=10$ GeV$^2$ (dotted line) as well as for $M_X^2=200$ GeV$^2$  at virtualities $Q^2=1$ GeV$^2$ (solid line), $Q^2=10$ GeV$^2$ (dashed  line) and $Q^2=50$ GeV$^2$ (dotted line).  These choices are based on the kinematic phase space for inclusive diffraction in $(x=\beta x_{\pom},Q^2)$ for the LHeC presented in Ref.  \cite{Armesto:2019gxy}. The location of the bump is strongly related to the value of the saturation scale and to the model we are using,  Eq. (\ref{qsnuc}), and the coefficient of proportionality between $(k_{\perp})_{max}$ and  $Q_{s,A}(x_{\pom})$  is equal to  $\kappa_{max}/Q_s \approx 1.5$ (we checked this is the case for any energy even at very low-$x_{\pom}$). That means the dimensionless cross section as a function of a scaling variable, $\tau_A=\kappa/Q_{s,A}$, is universal.  Just to exemplify quantitatively the value of the nuclear saturation scale in the domain considered above one has  $Q_{s,Pb}^2\approx 2.6$ GeV$^2$ (for $Q^2=1$ GeV$^2$ and $M_X^2=50$ GeV$^2$)   and  $Q_{s,Pb}^2\approx 1.7$ GeV$^2$    (for $Q^2=50$ GeV$^2$ and $M_X^2=200$ GeV$^2$) which are a factor 2 higher than in EIC case. This is translated into the jet transverse momentum at the peak, i.e., $(\kappa)_{max}\simeq 2.4$ GeV and $(\kappa)_{max}\simeq 2$ GeV, respectively. 
\begin{figure*}[t]
\subfloat[]{
\includegraphics[height=5.9cm]{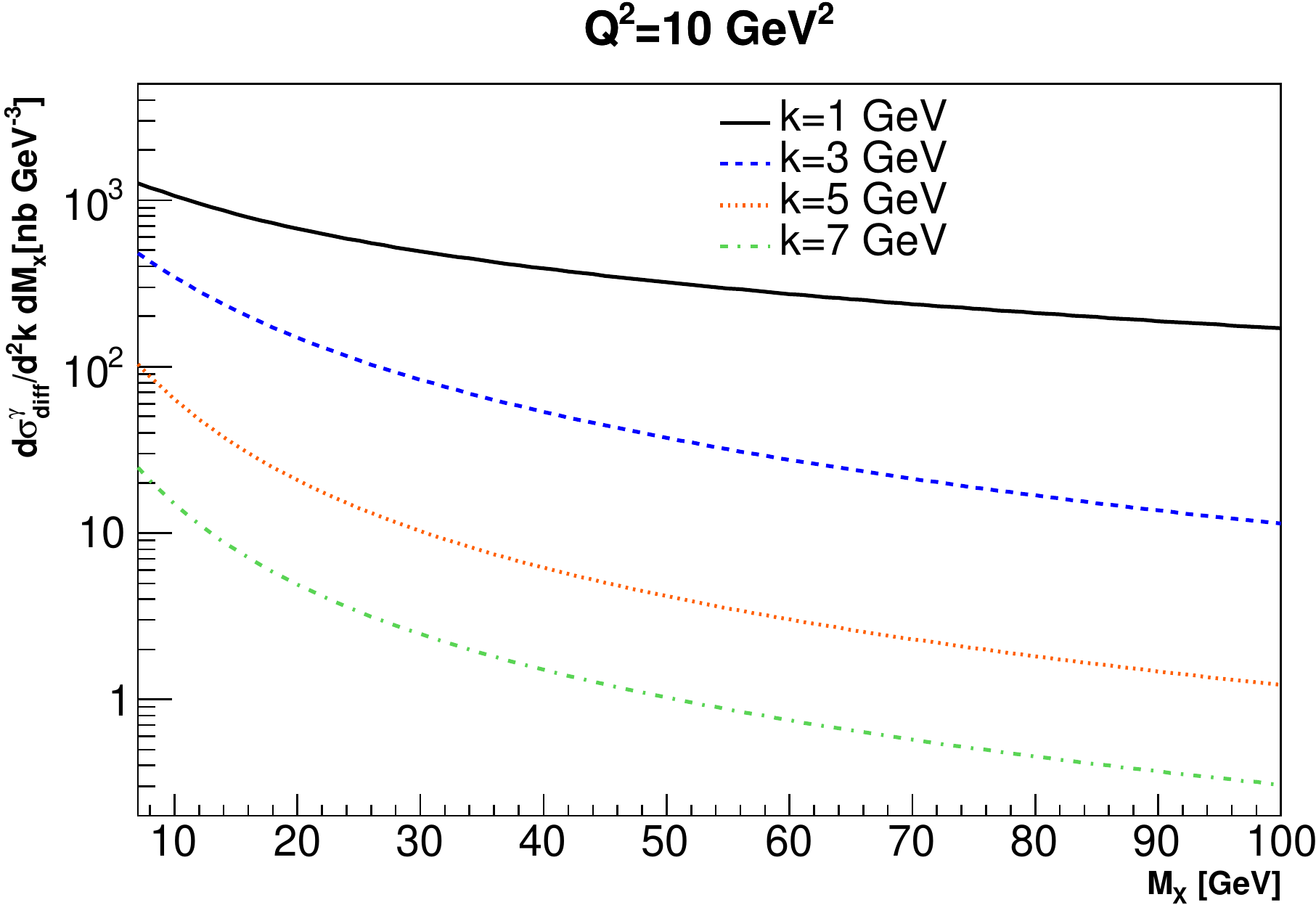}
\label{fig3a}
}
\quad %espaco separador
\subfloat[]{
\includegraphics[height=5.9cm]{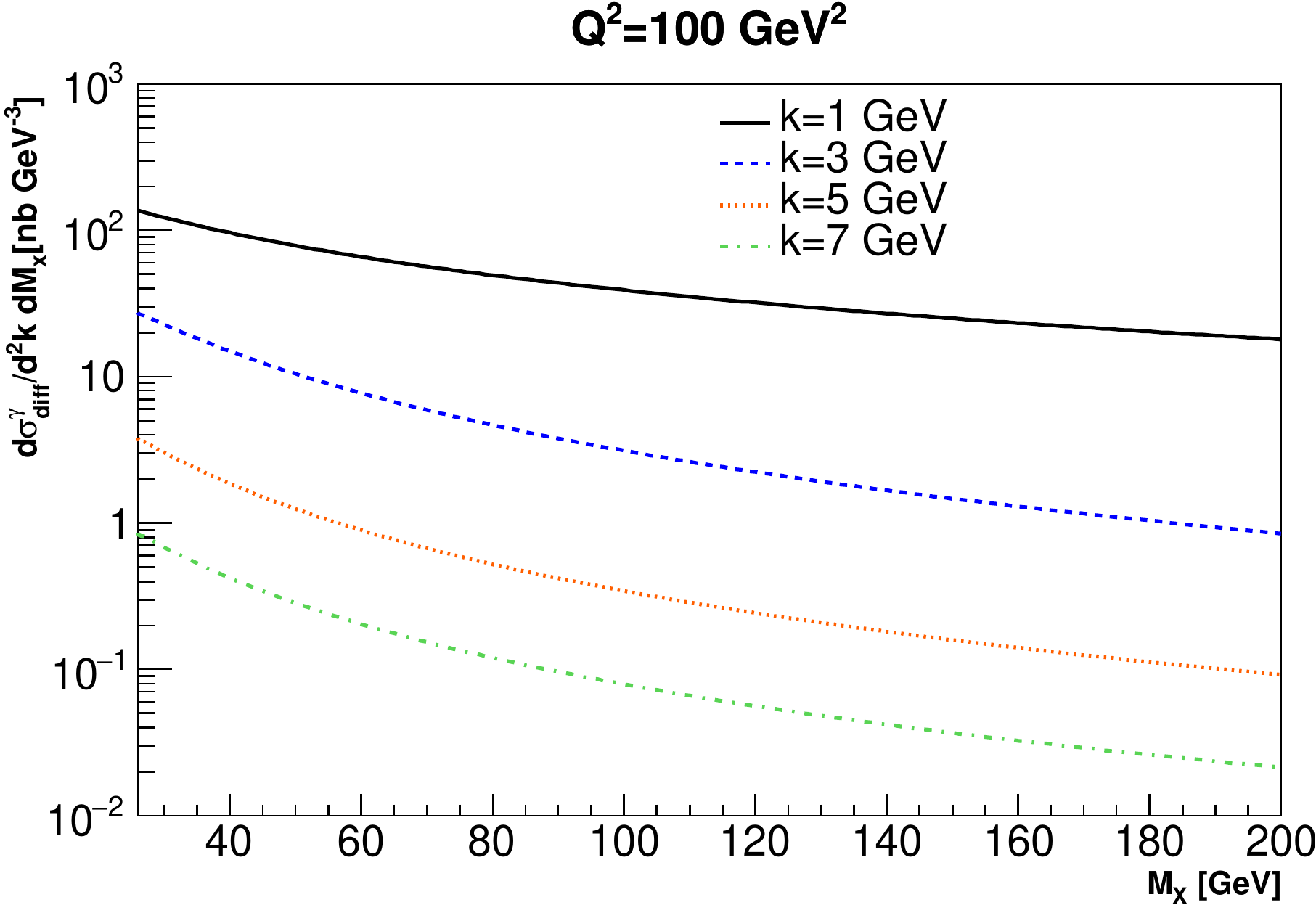}
\label{fig3b}
}
\caption{Differential cross section $d\sigma_{\mathrm{diff}}/d^2k_{\perp}dM_X$ as a function of diffractive mass $M_X$ for fixed $Q^2$ and $\kappa$ at the FCC-eA  ($\sqrt{s}=2.174$ TeV).  Two configurations are presented: (a)  $Q^2=10$ GeV$^2$ and  (b) $Q^2=50$ GeV$^2$. The jet transverse momentum increases in the curves from top to bottom.}
\label{fig:4}
\end{figure*}

Now we analyze the higher-energy upgrade of the LHeC (HE-LHeC) \cite{Bruning:2019scy,Bordry:2018gri} . The High-Energy Large Hadron Collider (HE-LHC) is a future energy upgrade of the LHC and its heavy-ion mode considers a  beam of nuclei with $E_A\simeq 4.9$ TeV per nucleon resulting in $\sqrt{s}\simeq 1.1$ TeV per nucleon. The expected luminosity is ${\cal{L}}=18\times 10^{32}$ cm$^{-2}$s$^{-1}$. In Fig. \ref{fig:3}, the cross section $\kappa^2 \sigma^{\mathrm{scaled}} (\kappa,Q^2,Q_{s,A})$ is plotted as a function of transverse momentum. We present the numerical results taking into account the same configuration as in the previous figure as a function of jet momentum. The general behavior remains the same, however the nuclear saturation scale has increased up to  $Q_{s,Pb}^2\approx 3$ GeV$^2$ and $Q_{s,Pb}^2\approx2 $ GeV$^2$   in the bins ($Q^2,\,M_X^2$) we had discussed before for LHeC.  The shift in the location of peak is now seen, where the maximum occurs for larger $\kappa$ in (a) compared to (b) due to the smaller $x_{\pom}$ value in that configuration. Accordingly, for the HE-LHeC the relation $(\kappa)_{max}\approx 1.5 \ Q_{s,A}$  still remains. As an example of numerical value of cross section, $M_Xd\sigma/d^2k_{\perp}dM_X\approx 7.4$ mb/GeV$^2$  at the peak for $Q^2=$ 1 GeV$^2$ and $M_X^2=200$ GeV$^2$.

\begin{figure}[t]
\includegraphics[width=1.0\columnwidth]{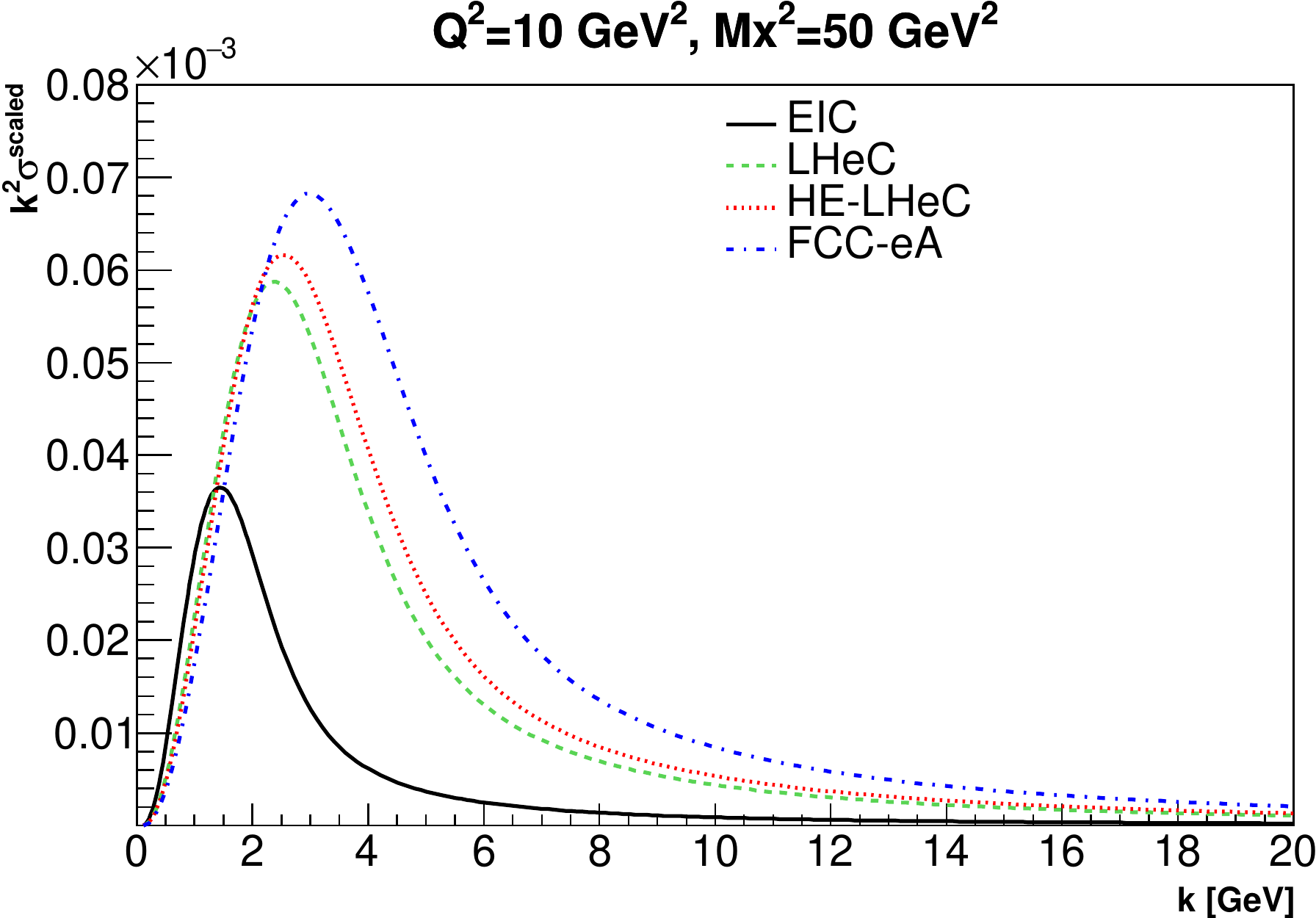}
\caption{The universal quantity $\kappa\sigma^{\mathrm{scaled}}$ as a function of jet transverse momentum for EIC, LHeC, HE-LHeC and FCC-eA machines for the sample configuration ($Q^2=10$ GeV$^2$, $M_X^2=50$ GeV$^2$). The displacement on the peak is proportional to the increasing nuclear saturation scale.}
\label{fig:5}
\end{figure}

Finally, we discuss $eA$ collisions at the FCC-eA \cite{Bruning:2019scy,Bordry:2018gri} machine that would be performed with a lead beam with energy per nucleon of $E_A=19.7$ TeV, which would give $\sqrt{s}\simeq 2.2$ TeV per nucleon with expected luminosity is ${\cal{L}}=54\times 10^{32}$ cm$^{-2}$s$^{-1}$. This is in the context of a Future Circular Collider - hadron-hadron mode (FCC-hh) that would provide $pp$ collisions with  $\sqrt{s}=100$ TeV. In Fig. \ref{fig:4}, the differential cross section $d\sigma/d^2k_{\perp}dM_X$ is presented as a function of $M_X$  for (a)  $Q^2=10$ GeV$^2$ (for fixed $\kappa=1, \ 3, \ 5, \ 7$ GeV) and (b) $Q^2=100$ GeV$^2$ (for fixed $\kappa=1,\,3,\,5,\,7$ GeV). In this figure, the jet transverse momentum increased in curves from top to bottom in panels (a) and (b). In Fig. \ref{fig:5}, we summarize the behavior of the scaled cross section times $\kappa^2$ for every collider and its machines as a function of jet momentum for the sample configuration $Q^2=10$ GeV$^2$ and $M_X^2=50$ GeV$^2$. The shift at the peak location is clearly seen, which is explained by the increasing of the nuclear saturation scale, $Q_{s,A}^2\approx 3 (x_0/x_{\pom})^{0.25}$ GeV$^2$, where $x_{\pom}\approx M_X^2/W_{\gamma p}^2$ in the region $\beta \ll 1$. All the findings we have discussed in $eA$ collisions should remain in $ep$ mode, where it is expected energies of order $\sqrt{s}=1-4$ TeV with luminosities ${\cal{L}}\simeq10^{34}$ cm$^{-2}$s$^{-1}$. In particular, $ep$ collisions at the LHeC can explore very low values of $\beta$ and a new domain of diffractive masses compared to DESY-HERA ($M_X$ can include $W/Z/$beauty or any state with $1^-$ quantum number).

%%%%%%%%%
\section{Summary and Conclusions}
\label{summary}

In this work, we have investigated the diffractive jet production in the small-$\beta$ region, which is dominated by large diffractive mass, $M_X^2\gg Q^2$. In the QCD color dipole picture, the main contribution comes from the $q\bar{q}g$ Fock state and the jet is associated to the soft gluon emitted. We study the potential of the  future EIC, LHeC and FCC-eA machines for the measurement of gluon jet diffractive cross section. In the TeV scale machines, one can reach $x_{\pom}\sim 10^{-5}$ for a wide range of $\beta$, corresponding to nuclear saturation scale of order $Q_{s,A}\simeq 2$ GeV. A simplified model for the $S$-matrices has been used and we discuss  the possible theoretical sources of uncertainty. As  examples of such sources  one has  more realistic expressions for the dipole-nucleus amplitude (Glauber model, Glauber-Gribov model or numerical solutions of BK equation) or different ansatz for the nuclear saturation scale.   Furthermore, we demonstrated that the nuclear saturation scale, $Q_{s,A}$, could be extracted from data as a function of $x_{\pom}$ by measuring the peak in the differential cross section $\kappa^2d\sigma/d^2k_{\perp}dM_X$ as a function of jet transverse momentum. Correlated strategies for extracting saturation scale from data are already known in literature. For instance, in Ref. \cite{Osada:2019oor} the proton saturation scale $Q_{s,p}^2$ is obtained from the multiplicities of charged hadrons in $pp$ collisions by using local parton-hadron duality and geometric scaling property (similar investigations were done for $pA$ \cite{McLerran:2015lta} and $AA$ collisions \cite{Andres:2012ma}). We present the probable region where the peaks occurs, $\kappa \approx a\times Q_{s,A}(x_{\pom})$ ($a$ is a constant of order of unity), and it was shown that the quantity $\kappa^2\sigma^{\mathrm{scaled}}$ presents universal behavior as a function of the scaling variable, $\tau=\kappa/Q_s$.  Summarizing, both the LHeC and its higher-energy version, the FCC-eh, offer unprecedented capabilities for studying the diffractive jet production in photon dissociation both in $ep$ and $ eA$ collisions.

\section*{Acknowledgments}

This work was financed by the Brazilian funding
agencies CNPq and  CAPES.

%\bibliographystyle{h-physrev}
%\bibliography{paper_eA}

\end{document}